\documentclass[epj]{svjour}
\usepackage{graphics}
\newcommand{\be}{\begin{equation}}
\newcommand{\ee}{\end{equation}}
\newcommand{\beq}{\begin{eqnarray}}
\newcommand{\eeq}{\end{eqnarray}}

\begin{document}
\title{Nonextensive statistical effects in protoneutron stars}
\author{A. Lavagno, D. Pigato}
\institute{Dipartimento di Fisica,
Politecnico di Torino, I-10129 Torino,
Italy and \\
Istituto Nazionale di Fisica Nucleare (INFN), Sezione di Torino, I-10126 Torino, Italy}
\date{Received: date / Revised version: date}
\abstract{We investigate the bulk properties of protoneutron stars in the framework of a
relativistic mean field theory based on nonextensive statistical
mechanics, characterized by power-law quantum distributions. We
study the relevance of nonextensive statistical effects on the
$\beta$-stable equation of state at fixed entropy per baryon, in presence and in absence of
trapped neutrinos, for nucleonic and hyperonic matter.
We show that nonextensive statistical effects could play a crucial role
in the structure and in the evolution of the protoneutron stars
also for small deviations from the standard Boltzmann-Gibbs
statistics.
 % end of PACS codes
} %end of abstract
\maketitle
\section{Introduction}\label{intro}
A protoneutron star (PNS) is born immediately after the
gravitational collapse of a massive star ($M\approx 10\div 20\,\,
M_{\odot}$). During the first seconds of its evolution a PNS is a
very hot, lepton rich and beta-stable object, with a tempe\-ra\-ture
of a few tens of MeV and a lepton concentration
typical of the pre-supernova matter \cite{BornofPNS}.
Nearly all of its binding energy is in the form
of neutrinos trapped inside the stellar
structure. In particular, neutrinos at the star core
are basically of the $\nu_e$ type and
have a typically energy of about $E_\nu
\approx 200\div300$ MeV \cite{cooke}. Shortly after the PNS formation,
during the cooling, neutrinos escape through the star structure
and bring out in few seconds nearly all of its binding energy. The
total luminosity of this process depends from several factors, the
most important are the total mass of the compact object and the
neutrino opa\-ci\-ty. During this process, called Kelvin-Helmholtz
epoch, the PNS evolves in a quasi hydrostatic equilibrium, from a
lepton rich and hot object to a cold neutrino-free compact star
\cite{PNS2}. From a measure of neutrinos luminosity and
average energy, which are the most important astrophysical
observable in the study of the stellar structure, it is in
principle possible to deduce the total binding energy and,
therefore, the total baryonic mass of the star \cite{PNSnero}.
However, neutrinos are not only important probes in the study of
newborn PNS, but have also relevant consequences on the chemical
composition and in the maximum mass of the compact object. The
presence of neutrinos in fact alters significantly the chemical
composition of the star, varying the formation threshold
of different particle species (in general the
appearance of non-leptonic negative particles is delayed when
neutrinos are present). The changes in the maximum
mass, due to neutrinos trapping, are normally greater than
those due to finite temperature effects \cite{cooke,PNS2,PNSnero,PNSrosso}.

From a more theoretical point of view, it appears evident that the
knowledge of the nuclear EOS plays a crucial role in the
determination of the structure and in the evolution of the PNS
\cite{noi_eos}. The processes related to strong interaction
should in principle be described by quantum chromodynamics.
However, in the energy density range reached in the compact stars,
strongly non-perturbative effects in the complex theory of QCD are
not negligible. The central core of a compact star does not at all
resemble a quasi-ideal gas of hadrons because strongly dynamical
correlations are present, including long-range interactions
\cite{unop,duep}. In the absence of a converging method to
approach QCD at finite density one often turns to (effective and phenomenological)
model investigations.

Recently, there is an increasing evidence that the nonextensive
statistical mechanics, characterized by power-law quantum
equilibrium distributions, can be considered as an appropriate
basis to deal with physical phenomena where strong dynamical
correlations, long-range interactions, anomalous diffusion and microscopic memory
effects take place \cite{tsallis,GMTsallis,book2,kodama}. A
considerable variety of physical applications involve a
quantitative agreement between expe\-ri\-mental data and theoretical
models based on the gene\-ra\-lized thermostatistics. In particular, in the
last years se\-ve\-ral authors have outlined the relevance of
nonextensive statistical mechanics effects in high energy physics
and astrophysical problems
\cite{bediaga,albe,beck,wilk1,plb2001,biroprl05,physicaA2008,cley,apj1,apj2,apj3,apj4,apj5}.
The existence of nonextensive statistical effects should strongly
affect the finite temperature and density nuclear EOS
\cite{pereira,andreada1}. In fact, by varying temperature and
density, the EOS reflects in terms of the macroscopic
thermodynamical variables the microscopic interactions in different nuclear matter regimes.

From a phenomenological point of view,
in this work we plan to investigate the relevance of nonextensive statistical
effects on the main physical properties of PNS and their related astrophysical implications.
At this scope, we are going to explore different stages of the PNS evolution.  The first stage corresponds to an entropy per baryon equal to one, in which neutrinos are trapped and strongly influence the chemical
composition of the PNS. After a short time, of about $10\div 15$
s, the temperature of the PNS fast rises up until it reaches a
value of $T\simeq 45\div 80$ MeV, it depends again on the chemical
composition \cite{Burrows}. This stage is called deleptonization
era and corresponds to the maximum heating and entropy per baryon
($s=2$). This is the phase, at high temperature and high baryon
density, in which the presence of nonextensive effects
may alter more sensibly the thermodynamical and mechanical
proprieties of the PNS. Furthermore, let us observe that we are
going to consider only hadronic composition of the PNS, and we
do not take in consideration possible formation of a mixed
hadron-quark phase region or a quark core inside it.
This choice is principally due to the fact
that the presence of trapped neutrinos delays the on-set of strange
particles and also possible formation of
quark matter \cite{cooke,PNS2}. As a consequence, the appearance
of a mixed phase or a quark core could be shifted at the end, or
nearly at the end of the deleptonization era \cite{lugones}.

The paper is organized as follows. In Section \ref{statistics}, we
briefly introduce the basic formalism of the nonextensive statistics. In
Section \ref{hadron}, we study the nonextensive hadronic EOS in
the context of a relativistic mean field model, in presence and in
absence of neutrinos and hyperons. In Section \ref{results}, we
show the main results in the particles concentrations and bulk
stellar properties in the presence of nonextensive statistical
effects. Finally, we summarize our conclusions in Section
\ref{conclusion}.

\section{Basic assumptions in nonextensive statistics}\label{statistics}

Starting point of the nonextensive statistical mechanics, introduced by Tsallis, is the following definition of a $q$-deformed entropy functional \cite{tsallis,GMTsallis,book2}:
\begin{eqnarray}
\!\!S_q[f]=\frac{1}{q-1} \left(1-\int[f({\bf x})]^q \,d\Omega\right), \left(\int f({\bf x})\,d\Omega=1\right)
\label{eq:GMTsallis}
\end{eqnarray}
where $f({\bf x})$ stands for a normalized probability
distribution, ${\bf x}$ and $d\Omega$ denoting, respectively, a
generic point and the volume element in the corresponding phase
space (here and in the following we set the Boltzmann and the
Planck constant equal to unity). The generalized entropy has the
usual properties of positivity, equiprobability, concavity and
irreversibility, preserves the whole mathematical structure of
thermodynamics (Legendre transformations). The real parameter $q$
determines the degree of non-additivity exhibited by the entropy
form (\ref{eq:GMTsallis}) and in the limit $q\rightarrow 1$,
becomes additive and reduces to the standard Boltzmann-Gibbs
entropy.

A peculiarity of the Tsallis generalized thermostati\-stics is that
if we have two statistically independent subsystems $A$ e $B$,
described, respectively, by individual proba\-bility density
$f^{(A)}$ and $f^{(B)}$ and we call $f^{(A+B)}({\bf x}_A,{\bf
x}_B)=$ $f^{(A)}({\bf x}_A)\,f^{(B)}({\bf x}_B)$ the joint
probability density of a composite system $A+B$, the nonadditive
(nonextensive) character of $S_q$ is summarized in the relation
\cite{book2}
\begin{eqnarray}
S_q[f^{(A+B)}]=&&S_q[f^{(A)}]+S_q[f^{(B)}]) \nonumber\\
&&+(1-q)\,S_q[f^{(A)}]\,S_q[f^{(B)}] \, . \label{next}
\end{eqnarray}
In the limit $q\rightarrow 1$, the above equation reduces to the
well-known additivity (extensivity) relation of the
Boltzmann-Gibbs logarithmic entropy. Here, the word nonextensive
should be associated with the fact that the total energy of
long-range-interacting mechanical systems is nonextensive, in
contrast with the case of short-range-interacting systems, whose
total energy is extensive in the thermodynamical sense
\cite{book2}.

Second crucial assumption on nonextensive statistics is the
introduction of the $q$-mean value (or escort mean value) of a
physical observable $A({\bf x})$ \cite{tsallis,GMTsallis,book2}:
\begin{equation}
 \displaystyle\langle A\rangle_q=\frac{\int A({\bf x})\,[f({\bf x})]^q
d\Omega} {\int [f({\bf x})]^q d\Omega} \, . \label{escort}
\end{equation}

The probability distribution can be obtained maximizing the measure
$S_q$ under appropriate constraints related to the previous
$q$-mean value definition. In this context, it is important
to observe that the Tsallis classical distribution can be seen as a
superposition of Boltzmann distributions with different temperatures
which have a mean value corresponding to the temperature appearing
in the Tsallis distribution. The nonextensive $q$ parameter is
related to the fluctuation in the temperature and describes the
spread around the average value of the Boltzmann temperature
\cite{wilk1}.

Following the above prescriptions, it is possible to obtain the
associate quantum mean occupation number of particles species $i$
in a grand canonical ensemble. For a dilute gas of particles and
for small deviations from the standard statistics ($q\approx 1$),
it can be written as \cite{tewe,NJL}
\begin{equation}
n_i=\frac{1} { {\tilde{e}}_q(\beta(E_i-\mu_i)) \pm 1} \, ,
\label{eq:distribuzioneq>1a}
\end{equation}
where $\beta=1/T$ and the sign $(\pm 1)$ is for fermions and
bosons respectively. Furthermore, in
Eq.(\ref{eq:distribuzioneq>1a}), following Ref. \cite{NJL}, for $q>1$, we have
\begin{equation}
{\tilde{e}}_q(x)=\cases{ [1+(q-1)x]^{1/(q-1)} &if $x>0$ ; \cr
[1+(1-q)x]^{1/(1-q)} &if $x\le 0$ ,\cr} \label{eq:distribuzioneq2}
\end{equation}
whereas, for $q<1$,
\begin{equation}
{\tilde{e}}_q(x)=\cases{ [1+(q-1)x]^{1/(q-1)} &if $x\le 0$ ; \cr
[1+(1-q)x]^{1/(1-q)} &if $x>0$ .\cr} \label{eq:distribuzioneq3}
\end{equation}
Naturally, for $q\rightarrow1$ the above quantum distribution
reduces to the standard Fermi-Dirac and Bose-Einstein
distribution. Let us observe nonextensive statistical effects
vanishes approaching to zero temperature. On the other hand,
the nonextensive statistics entails a sensible difference on the
power-law particle distribution shape in the high energy region
with respect to the standard statistics. Hence, nonextensive
effects could play an important role in the finite
temperature and high baryon density PNS evolution.

\section{Nonextensive hadronic equation of state}
\label{hadron}

In this Section, we study the nonextensive hadronic EOS in the
framework of a relativistic mean field theory in which baryons
interact through the nuclear force mediated by the exchange of
virtual isoscalar-scalar ($\sigma$), isosca\-lar-vector ($\omega$)
and isovector-vector ($\rho$) mesons fields
\cite{walecka,boguta,glen}.
In our analysis we include all the baryon octet
in order to reproduce the chemical composition of the PNS
at high baryon chemical potential. We also take into account of
leptons particle by fixing the lepton fraction
$Y_{L}=Y_e+Y_{\nu_e}=(\rho_{e}+\rho_{\nu_e})/\rho_{B}$, where
$\rho_e$, $\rho_{\nu_e}$ and $\rho_B$ are the electron, neutrino
and baryon number densities, respectively. This is because, in the
first stage of PNS evolution, electrons and neutrinos are trapped
inside the stellar matter and, therefore, the lepton number must be
conserved until neutrinos escape out of the PNS
\cite{cooke,PNSnero,PNSrosso}.

The Lagrangian density can be written in term of the hadronic
\cite{glen,ditoro} plus leptonic component, as follow:
\begin{eqnarray}\label{eq:111}
&&{\mathcal L}_{tot}={\mathcal
L}_{H}+{\mathcal L}_{l}=
\sum_{B}\bar{\psi}_B
[i\gamma_{\mu}\partial^{\mu}-(M_B- g_{\sigma B}\sigma) \nonumber\\
&&-g{_\omega}_{B}\gamma_\mu\omega^{\mu}-g_{\rho B}
\gamma^{\mu}\vec\tau\cdot
\vec{\rho}_{\mu}]\psi_B
+\frac{1}{2}(\partial_{\mu}\sigma\partial^{\mu}\sigma-m_{\sigma}^2\sigma^2)\nonumber \\
&&-U(\sigma)+\frac{1}{2}m^2_{\omega}\omega_{\mu}
\omega^{\mu}
+\frac{1}{2}m^2_{\rho}\vec{\rho}_{\mu}\cdot\vec{\rho}^{\;\mu}
-\frac{1}{4}F_{\mu\nu}F^{\mu\nu}\nonumber \\
&&-\frac{1}{4}\vec{G}_{\mu\nu}\vec{G}^{\mu\nu}\ +\sum_{l}\bar{\psi}_l
[i\gamma_{\mu}\partial^{\mu}-m_l]\psi_l \, ,
\end{eqnarray}
where the sums over $B$ and $l$ are over the baryon octet and
lepton particles, respectively. The field strength tensors for the vector mesons are
given by the usual expressions
$F_{\mu\nu}\equiv\partial_{\mu}\omega_{\nu}-\partial_{\nu}\omega_{\mu}$,
$\vec{G}_{\mu\nu}\equiv\partial_{\mu}\vec{\rho}_{\nu}-\partial_{\nu}\vec{\rho}_{\mu}$,
and $U(\sigma)$ is a nonlinear potential of $\sigma$ meson
\begin{eqnarray}
U(\sigma)=\frac{1}{3}a\sigma^{3}+\frac{1}{4}b\sigma^{4}\,,
\end{eqnarray}
usually introduced to achieve a reasonable compression modulus for
equilibrium nuclear matter.

The field equations in a mean field approximation are
\begin{eqnarray}
&&\!\!\!\!\!\!\!(i\gamma_{\mu}\partial^{\mu}-(M- g_{\sigma B}\sigma)-
g_{\omega B}\gamma^{0}\omega-g_{\rho B}\gamma^{0}{\tau_3}\rho)\psi=0\,, \\
&&\!\!\!\!\!\!\!m_{\sigma}^2\sigma+ a{{\sigma}^2}+ b{{\sigma}^3}=
g_{\sigma B}<\bar\psi\psi>=g_{\sigma B}{\rho}_S\,, \\
&&\!\!\!\!\!\!\!m^2_{\omega}\omega=g_{\omega B}<\bar\psi{\gamma^0}\psi>=g_{\omega B}\rho_B\,,\\
&&\!\!\!\!\!\!\!m^2_{\rho}\rho=g_{\rho B} <\bar\psi{\gamma^0}\tau_3\psi>=g_{\rho
B}\rho_I\,, \label{eq:MFT}
\end{eqnarray}
where $\sigma=\langle\sigma\rangle$, $\omega=\langle\omega^0\rangle$
and $\rho=\langle\rho^0_3\rangle$ are the nonvanishing expectation
values of mesons fields, $\rho_I$ is the total isospin density,
$\rho_B$ and $\rho_S$ are the baryon density and the baryon scalar
density, respectively. They are given by
\begin{eqnarray}
&&\rho_{B}=2 \sum_{i=B} \int\frac{{\rm
d}^3k}{(2\pi)^3}[n_i(k)-\overline{n}_i(k)]\,, \label{eq:rhob} \\
&&\rho_S=2 \sum_{i=B} \int\frac{{\rm
d}^3k}{(2\pi)^3}\,\frac{M_i^*}{E_i^*}\,
[n_i^q(k)+\overline{n}_i^{\,q}(k)]\,, \label{eq:rhos}
\end{eqnarray}
where $n_i(k)$ and $\overline{n}_i(k)$ are the $q$-deformed
fermion particle and antiparticle distributions given in
Eq.s(\ref{eq:distribuzioneq>1a})-(\ref{eq:distribuzioneq3}); for
example, for $q>1$ and $\beta(E_i^*-\vert\mu_i^*\vert)>0$, we have
\begin{eqnarray}
n_i(k)=\frac{1} { [1+(q-1)\,\beta(E_i^*(k)-\mu_i^*)
]^{1/(q-1)} + 1} \label{eq:distribuz} \, , \\
\overline{n}_i(k)=\frac{1} {[1+(q-1)\,\beta(E_i^*(k)+\mu_i^*)
]^{1/(q-1)} + 1} \, . \label{eq:distribuz2}
\end{eqnarray}

The nucleon effective energy is defined as
${E_i}^*(k)=\sqrt{k^2+{{M_i}^*}^2}$, where ${M_i}^*=M_{i}-g_{\sigma
B} \sigma$. The effective chemical potentials $\mu_i^*$  are given
in terms of the meson fields as follows
\begin{eqnarray}
\mu_i^*={\mu_i}-g_{\omega B}\omega -\tau_{3i B} g_{\rho B}\rho \, ,
\label{mueff}
\end{eqnarray}
where $\mu_i$ are the thermodynamical chemical potentials
($\mu_i=\partial\epsilon/\partial\rho_i$).

The further conditions, required for the $\beta$-stable chemical
equilibrium and charge neutrality, can be written as
\begin{eqnarray}
&&\mu_{\Lambda}=\mu_{\Sigma^{0}}=\mu_{\Xi^{0}}=\mu_{n}\, ,\\
&&\mu_{\Sigma^{-}}=\mu_{\Xi^{-}}=\mu_n+ \mu_e\, ,\\
&&\mu_p=\mu_{\Sigma^{+}}=\mu_n-\mu_e \, ; \\
&&\rho_{p} + \rho_{\Sigma^{+}} - \rho_{\Sigma^{-}} -
\rho_{\Xi^{-}} - \rho_{e}=0 \, .
\end{eqnarray}
In the case of trapped neutrinos, the new equalities  are obtained
by the replacement of $\mu_e\rightarrow\mu_e-\mu_{\nu_{e}}$.  The
total entropy per baryon is calculated using
$s=(S_B+S_l)/(T\rho_B)$, where $S_B=P_B+\epsilon_B -
\sum_{i=B}\mu_i\rho_i$ and $S_l=P_l+\epsilon_l-
\sum_{i=l}\mu_i\rho_i$, and the sums are extended over all the
baryons and leptons species.

The thermodynamical quantities can be obtained from the
thermodynamic potential in the standard way. More explicitly, the
baryon pressure $P_B$ and the energy density $\epsilon_B$ can be
written as
\begin{eqnarray}
P_B=&&\frac{2}{3} \sum_{i} \int \frac{{\rm d}^3k}{(2\pi)^3}
\frac{k^2}{E_{i}^*(k)} [n_i^q(k)+\overline{n}_i^q(k)]
-\frac{1}{2}m_\sigma^2\sigma^2  \nonumber \\
&&- U(\sigma)+\frac{1}{2}m_\omega^2\omega^2+\frac{1}{2}m_{\rho}^2 \rho^2\,,\label{eq:eos}\\
\epsilon_B=&& 2 \sum_{i}\int \frac{{\rm d}^3k}{(2\pi)^3}E_{i}^*(k)
[n_i^q(k)+\overline{n}_i^q(k)]
+\frac{1}{2}m_\sigma^2\sigma^2 \nonumber \\
&&+U(\sigma)+\frac{1}{2}m_\omega^2\omega^2+\frac{1}{2}m_{\rho}^2 \rho^2\, .
\label{eq:eos2}
\end{eqnarray}
The numerical evaluation of the above thermodynamical quantities
can be performed if the meson-nucleon and meson-hyperon coupling
constants are known. Concerning the meson-nucleon coupling
constants ($g_{\sigma N}$, $g_{\omega N}$, $g_{\rho N}$), they are
determined to reproduce properties of equilibrium nuclear matter
such as the saturation densities, the binding energy, the
symmetric energy coefficient, the compression modulus, and the
effective Dirac mass at saturation. Because of a valuable range of
uncertainty in the empirical values that must be fitted,
especially for the compression modulus and for the effective Dirac
mass, in the literature there are different sets of coupling
constants. Here and in the following, we focus our investigation
by considering the so-called GM3 parameter set (even if comparable results
can be obtained in other parameter sets) \cite{glen,ditoro}. The
implementation of hyperon degrees of freedom comes from
determination of the corresponding meson-hyperon coupling
constants that have been fitted to hypernuclear properties and
their specific values are taken from Ref.\cite{ANDREA} for the GM3
parameter set.

\section{Results and discussion}
\label{results}

As briefly mentioned in the Introduction, our analysis focus on the relevance of
possible nonextensive statistical effects during the first PNS evolution phases.
We can ideally divide the evolution into three
phases. The first, at the beginning, in which neutrinos are
trapped and the entropy per baryon is assumed fixed to $s=1$ and
$Y_{L}=0.4$. A second phase, after about $10\div 15$ sec, which
corresponds to the maximum heating of the star and neutrinos
are free ($s=2, Y_{\nu_e}=0$). Finally, a third phase of
cold-catalyzed PNS ($s=0, Y_{\nu_e}=0$) \cite{Burrows,yle}.
Regarding the relevance of nonextensive statistical effects,
the most important phase corresponds to the maximum heating,
in which the presence of nonextensive statistical effects may
play a crucial role in the determination of the PNS chemical
composition and related thermodynamical proprieties. In this work
we limit ourselves to consider only a small variations from the
standard statistics (from $q=0.97$ to $q=1.03$).

\begin{figure}
\begin{center}
\resizebox{0.48
\textwidth}{!}{%
\includegraphics{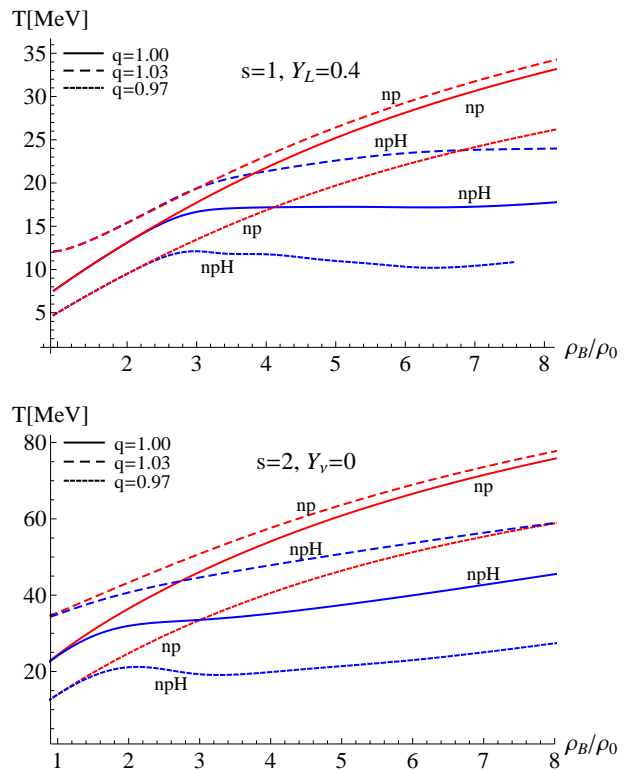}
} \caption{Temperature as a function of the baryon density (in
units of the saturation nuclear density $\rho_0$) for different values of $q$, entropy per baryon
and neutrino fraction (upper panel: $s=1, Y_{L}=0.4$ and lower panel: $s=2, Y_{\nu_e}=0$). The labels $np$ and $npH$ stand for
nucleons and nucleons plus hyperons.} \label{fig:temp1}
\end{center}
\end{figure}

In Fig. \ref{fig:temp1}, we show the temperature
as a function of the baryon density (in units of the saturation nuclear density $\rho_0=0.153$
fm$^{-3}$) and for different values of the nonextensive parameter, in absence ($np$) and in presence ($npH$) of hyperons. We limit our analysis in the first two
phases: in the upper panel, the first leptonic rich state ($s=1, Y_{L}=0.4$) and,
in the lower panel, the maximum heating phase ($s=2, Y_{\nu_e}=0$).
Indeed in the cold-catalyzed phase ($s=0,
Y_{\nu_e}=0$), the temperature is very low (fews MeV), and
nonextensive statistical effects may be neglected. In both previous cases, we
observe a reduction in temperature in presence of a
sub-extensive stati\-stics ($q<1$) and a general increase for $q>1$.
This effect is more remarkable when hyperons are present and
for higher values of entropy for baryon.
For example, in the maximum heating phase, in presence of
hyperons and for $q=1$, the temperature is about $T\cong 37$
MeV at baryon density $\rho_B= 5\rho_0$, whereas, for $q=1.03$ and $q=0.97$, it is approximately equal to $T\cong 51$ MeV and $T\cong 21$ MeV, respectively.
Note also that, when hyperons are present, for $s=1$ and
$Y_{L}=0.4$, the system evolves in a quasi
isothermal configuration above $\rho_B= 2.5\div 3\,\rho_0$.
The different behavior in the stellar temperature have important consequences
in the PNS evolution and in its particles concentration.
Finite temperature properties of matter at high density influence the diffusion of neutrinos,
being the neutrino mean free paths strongly temperature dependent \cite{PNSnero,yle}.
In particular, neutrino opacity is very
sensitive to the inner temperature (in general proportional to
$T^2$) and, therefore, this would affect sensibly the
cooling of the PNS, making it longer when $q>1$, and shorter when $q<1$.
This matter of fact could have important consequences on the neutrino luminosity,
because its drop is associated with the end of this cooling process
\cite{PNS2}. Consequently, an alteration in the Kelvin-Helmholtz
epoch, that does not correspond to the predict direct or modified
URCA process, could be an indication of nonextensive statistical effects.

\begin{figure}
\begin{center}
\resizebox{0.48\textwidth}{!}{%
\includegraphics{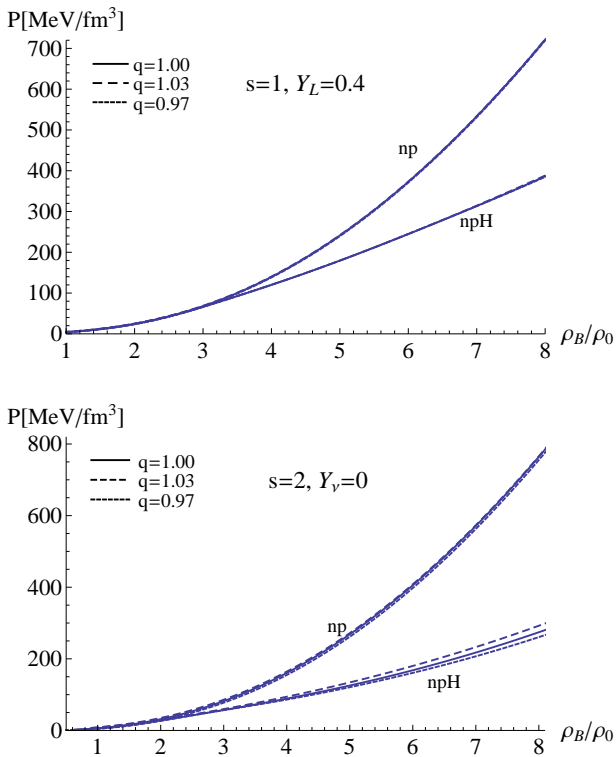}
} \caption{Total pressure as a function of baryon density (in units of the saturation nuclear density) for different values of $q$ in the lepton rich case: $s=1$ and
$Y_{L}=0.4$ (upper panel) and in the maximum heating phase: $s=2$ and
$Y_{\nu_e}=0$ (lower panel). The labels $np$ and $npH$ stand for
nucleons and nucleons plus hyperons matter, respectively.}
\label{fig:Prhob1}
\end{center}
\end{figure}

In Fig. \ref{fig:Prhob1}, is reported the dependence of the pressure
from the baryon density, for different values of $q$, in the initial
phase: $s=1, Y_{L}=0.4$ (upper panel) and in the
maximum heating phase: $s=2$ and $Y_{\nu_e}=0$ (lower panel). With
the appearance of hyperons, around $\rho_B= 3\,\rho_0$, we have
a general softening of the EOS. However, due to the low temperature
achieved in this phase (see Fig. \ref{fig:temp1}), nonextensive statistical
effects do not change significantly the total pressure of the PNS.
The situation is somewhat different when
we analyze the maximum heating phase. In such a condition, the
temperature is higher and nonextensive statistical effects are more
relevant, especially when hyperons are present.

\begin{figure}
\begin{center}
\resizebox{0.48\textwidth}{!}{%
\includegraphics{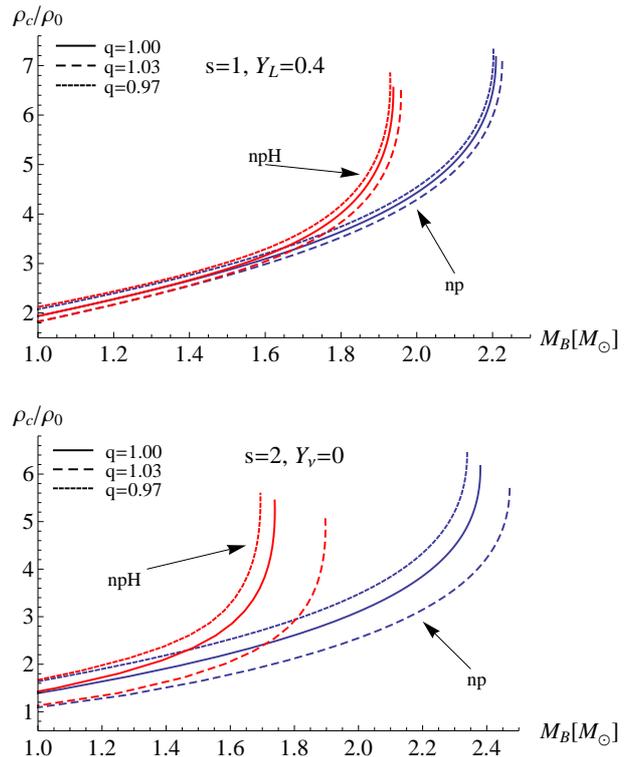}
} \caption{Central baryon density $\rho_c$ (in units of the saturation nuclear density) corresponding to a total stellar baryon mass $M_B$ in the lepton rich phase (upper panel) and in the maximum heating phase (lower panel). The labels $np$ and $npH$ stand for nucleons and nucleons plus hyperons matter, respectively.}
\label{fig:rhoc}
\end{center}
\end{figure}

As it is well known, the softening of the EOS, due to the appearance of additional fermionic degrees of freedom in the form of hyperons, leads to higher central densities. This matter of fact is,  however, influenced from the presence of nonextensive statistical effects. To better focalized this aspect, in Fig. \ref{fig:rhoc}, is plotted the central baryon density $\rho_c$ in the PNS core corresponding to a total baryon mass $M_B$ (in units of the solar mass $M_\odot$). The case $q>1$ ($q<1$) implies a general reduction (increase) of the central density at fixed total baryon mass. This effect is emphasized in the maximum heating condition (lower panel). In particular, for $q>1$ and until $M_B\approx 1.7\, M_\odot$, the central densities of hyperons stars ($npH$) are lower than the ones corresponding to the standard ($q=1$) nucleons-only stars ($np$). This feature, together the variation of the temperature, can be very relevant in the determination of the specific heat of the stellar matter and, as a consequence, on the neutrino diffusion \cite{PNSnero}.

\begin{figure}
\begin{center}
\resizebox{0.48\textwidth}{!}{%
\includegraphics{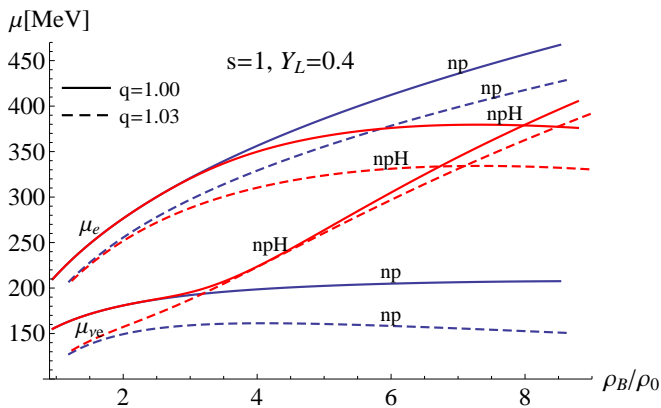}
} \caption{Lepton chemical potentials in lepton rich matter as a function of the baryon density (in units of the nuclear saturation density). The labels $np$ and $npH$ stand for nucleons and nucleons plus hyperons matter, respectively.}
\label{fig:mu}
\end{center}
\end{figure}

In addition to the above considerations, we investigate the lepton chemical potentials in different PNS conditions because they strongly influence the deleptonization process \cite{PNSnero,PNSrosso}. At this scope, in Fig. \ref{fig:mu}, we show the lepton chemical potentials in lepton rich matter ($s=1$ and $Y_{L}=0.4$) in absence ($np$) and in presence ($npH$) of hyperon degrees of freedom (higher curves for the electron chemical potential $\mu_e$ and lower curves for the neutrino chemical potential $\mu_{\nu_e}$). For clarify, we report only the case $q>1$ compared to the standard ($q=1$) results. It is interesting to observe that for a nucleons-only EOS, super-extensive statistical effects imply a general reduction of the electron and neutrino chemical potentials respect to the standard case. Otherwise, in presence of hyperons, a sensible reduction of the electron chemical potential is not accompanied by a reduction of the neutrino chemical potential which retains very similar values to the standard case at high baryon density. Such a feature could have important consequence on the diffusion of the electron neutrinos inside the PNS.

\begin{figure}
\begin{center}
\resizebox{0.48
\textwidth}{!}{%
\includegraphics{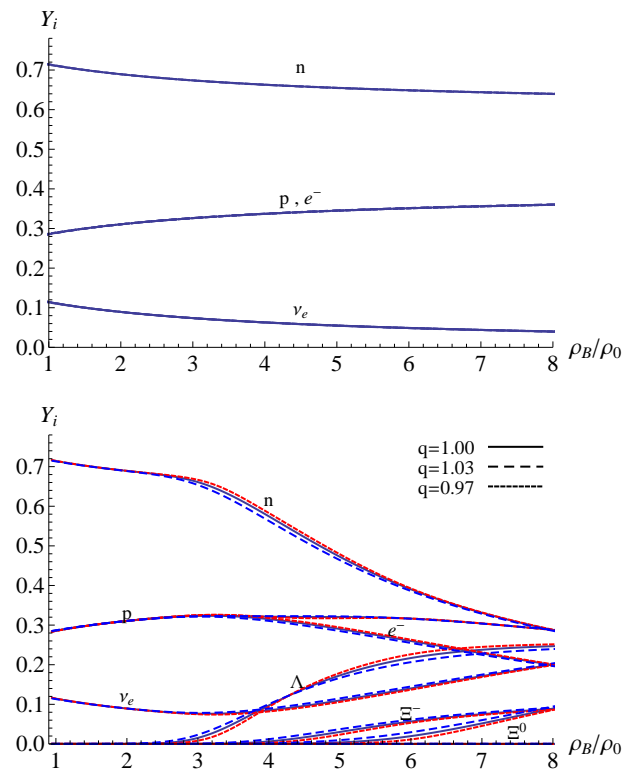}
} \caption{Particle concentrations $Y_i$ without (upper panel) and
with hyperons (lower panel) as a function of the baryon density for
$s=1$ and $Y_{L}=0.4$.} \label{fig:concnp1}
\end{center}
\end{figure}

Moreover, the neutrino mean free paths and the matter specific heat depend sensitively on the composition; under degenerate conditions even modest changes to the composition significantly alter the neutrino scattering and absorbtion mean free paths. It is, therefore, relevant to investigate how nonextensive statistical mechanics influences particle compositions in different PNS epochs.
In Fig. \ref{fig:concnp1}, we report the particle concentrations for
$s=1$ and $Y_{L}=0.4$ in absence (upper panel) and in presence (lower panel) of hyperons for different values
of $q$. It is well known that the presence of trapped neutrinos
significantly alter the protons and the electrons abundance and
strongly influence the threshold of hyperons formation \cite{cooke,PNSnero,PNSrosso}.
In absence of hyperons, nonextensive statistical effects do not play a significantly role, in fact the particle concentrations are almost the same. The situation changes when we include
hyperon degrees of freedom.
The presence of sub-extensive effects ($q<1$) slightly lowers the neutrinos
concentration, while increases the neutrons and the electrons ones.
Moreover, hyperons start later and their concentration are in
general decreased, excepted for the $\Lambda$ particles ratio, which becomes greater
after $\rho_B> 4\,\rho_0$ respect to the standard case. Otherwise, in case of super-extensive
statistical effects ($q>1$), we can observe a general reduction in the neutrons
and the electrons fractions and a small increase of the neutrinos and
the hyperons concentrations, with important effects on the softening of the EOS.

\begin{figure}
\begin{center}
\resizebox{0.48
\textwidth}{!}{%
\includegraphics{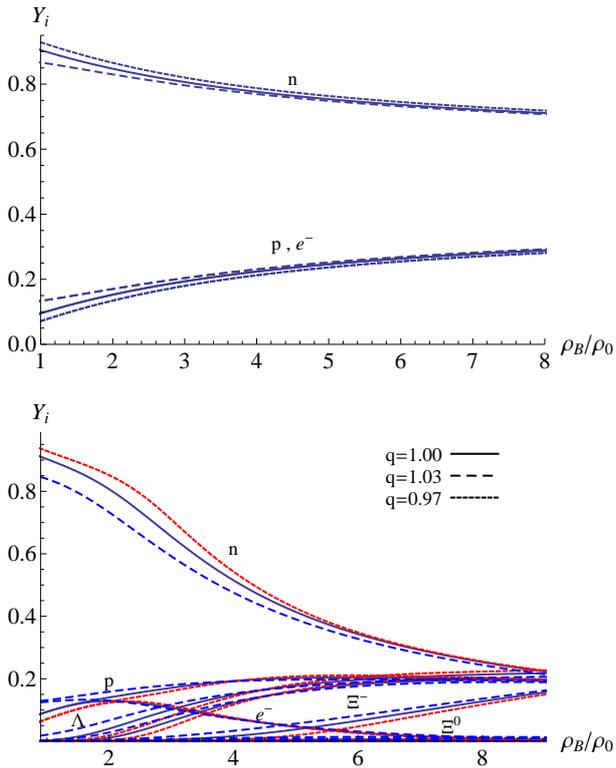}
} \caption{The same of Fig. \ref{fig:concnp1} for the maximum
heating phase ($s=2$ and $Y_{\nu_e}=0$).} \label{fig:concnp2}
\end{center}
\end{figure}

In the Fig. \ref{fig:concnp2}, particle concentrations for the maxi\-mum heating phase
($s=2$, $Y_{\nu_e}=0$) are reported. In this condition, due to the higher
temperature achieved in the stellar matter, nonextensive statistical effects
become more relevant and, consequently, particle concentrations
change significantly. When $q<1$, we observe a reduction in
the protons and the electrons concentrations and an increase of the
neutrons fraction. Whereas, in the super-extensive case ($q>1$), we have a
lower neutrons fraction and an increase of the protons and
electrons concentrations.
When hyperons are included, Fig. \ref{fig:concnp2}, lower panel,
we have two main consequences.
Firstly, with the absence of neutrinos, the hyperons on-set is shifted
at low baryon densities, below $2 \,\rho_0$. Secondly, as a consequence,
the electrons and the protons concentrations decrease sensibly with respect to the initial lepton-rich regime ($s=1, Y_{L}=0.4$). Therefore, the
absence of neutrinos in the stellar matter implies a strong
softening of the EOS. In presence of sub-extensive statistics ($q<1$),
hyperons start later with respect to the standard case ($q=1$), and have in
general a bigger concentration at high baryon density. The other way round takes place
for super-extensive statistics ($q>1$).

\begin{figure}
\begin{center}
\resizebox{0.48
\textwidth}{!}{%
\includegraphics{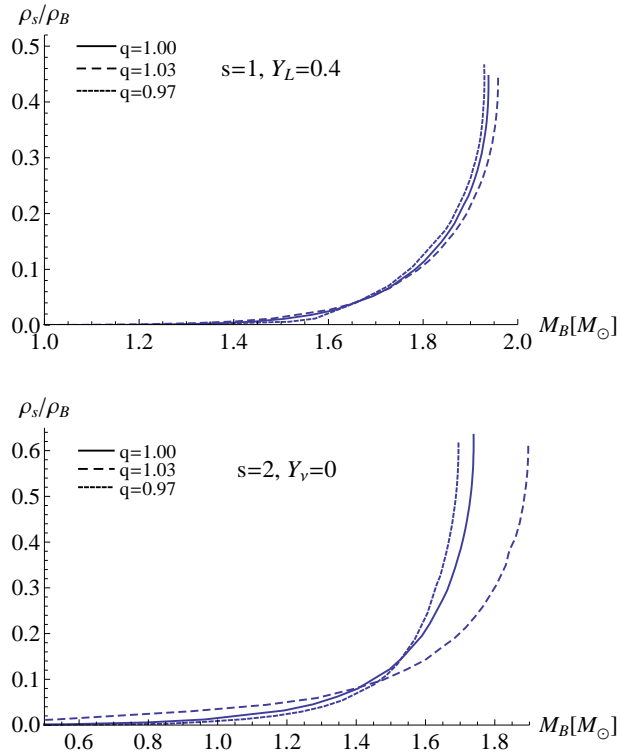}
} \caption{Hyperons concentration $\rho_s/\rho_B$ (strangeness per baryon) as a function of the baryon mass $M_B$ in the lepton rich (upper panel) and in the maximum heating phase (lower panel). } \label{fig:rhos}
\end{center}
\end{figure}

To better understand the role of nonextensive statistical effects on hyperons formation in the PNS core, in Fig. \ref{fig:rhos}, we report the total hyperons concentration (strange\-ness per baryon) as a function of the stellar baryon mass (in units of $M_\odot$) in the lepton rich (upper  panel) and in the maximum heating (lower panel) epoch. Although, for $q>1$, hyperons are present at lower baryon densities with respect to the standard case ($q=1$) and a greater hyperons concentration at lower baryon masses is present, we have a significant reduction at higher baryon masses, especially in case of the maximum heating condition. For $q<1$, we have instead the opposite effect: a reduction of the hyperons fraction at low baryon masses and an enhancement at high baryon masses. These features, principally due to the behavior of $\Lambda$ particles concentration as a function of the baryon density (see Fig.s \ref{fig:concnp1} and \ref{fig:concnp2}), could imply relevant phenomenological consequences on the evolution of the PNS. In fact, it is known that hyperons significantly increase the neutrino scattering and absorption cross sections \cite{PNSnero}. Furthermore, the central densities of hyperons stars become progressively larger than that of purely nucleon stars and the evolution timescale of hyperons stars results to be  slightly larger because of the smaller mean free path of hyperonic matter. Larger central densities and higher electron neutrino energies,  reached in hyperonic PNS, increase the neutrino opacity, temporarily reducing the loss of neutrinos from the stellar core and allow to sustain a higher luminosity at late times.

Finally, in the Table \ref{tab:masse}, we report the maximum gravitational (baryonic) masses (in units of $M_\odot$) and the corresponding values of radius
and central baryon density, for different PNS configurations. We have taken in consi\-de\-ration different values of the nonextensive $q$-parameter, entropy per baryon and leptons
concentration, in absence ($np$) and in presence ($npH$) of hyperons.

As we have already remarked, nonextensive statistical effects result to be
most relevant when hyperons are present and the maximum heating phase is achieved
($s=2$, $Y_{\nu_e}=0$). In fact, when the entropy per baryon is equal to one and the matter
is lepton rich, the difference in the maximum mass is very small both with and without hyperons.
Moreover, immediately after the deleptonization, nonextensive
statistical effects become more important, especially in presence of hyperons.
The maximum mass is in general decreased when we consider a sub-extensive statistics ($q<1$),
and increased when $q>1$.

Furthermore, in the presence of hyperons, when the stellar core contains non-leptonic negative
charges, the maximum masses of neutrino-trapped stars result to be significantly larger
than for low temperatures and for lepton poor matter. Hence, there exists a window of initial
masses for which the star becomes unstable to gravitational collapse during deleptonization
and a black hole can take place \cite{cooke,PNSnero,PNSrosso}.
We can see that the formation of such a metastable phase strongly depends on the presence
of nonextensive statistical effects and the window of metastability grows with the value of the nonextensive index
$q$. In particular, for $q>1$, we still have a very large value of the maximum baryonic mass in
the maximum heating phase ($s=2$, $Y_{\nu_e}=0$), significantly larger than the one in the cold catalyzed condition
(where nonextensive statistical effects do not play any role).
Therefore, for nucleons-only stars, a black hole could form only
after the core bounce, because the maximum mass supported by
neutrino-free stars is bigger than that supported for neutrino-rich
case. Whereas, in presence of hyperons, a black hole can take place also after the deleptonization era because of the realization of a metastable phase, which becomes more relevant in presence of super-extensive statistical effects.

\begin{table}
\caption{Maximum gravitational (baryonic) masses $M_{\rm max}$ (in units of $M_\odot$) and  corresponding values of radius $R$ and central baryon
density $\rho_c$ in absence ($np$) and in presence ($npH$) of hyperons
for different values of the nonextensive parameter $q$. The
results are reported for different values of entropy per baryon and leptons concentration. For completeness, the values for the cold-catalyzed phase ($s=0, Y_{\nu_e}=0$) for $q=1$ are also reported (nonextensive statistical mechanics does not play any role in this last regime).}
\label{tab:masse}
%\vspace{0.5cm}
\begin{tabular}{|c|c|c|}
\hline \hline
    & \multicolumn{2}{|c|}{}                                        \\
    & \multicolumn{2}{|c|}{$np$}                                 \\
    & \multicolumn{2}{|c|}{}                                       \\
\hline \hline
%    &                   &                                        \\
$q$ &$s=1$, $Y_{L}=0.4$ & $s=2$, $Y_{\nu_e}=0$                   \\
%    &                   &                                        \\
\hline \hline
%    &                   &                                        \\
     &$M_{\rm max}=1.96\, (2.20)$  & $M_{\rm max}=2.06\, (2.34)$ \\
0.97 &R=10.52 km                   & R=11.37 km                  \\
     &$\rho_c=7.33 \,\rho_0$       & $\rho_c=6.50 \,\rho_0$      \\
%    &                   &                                        \\
\hline
%    &                   &                                        \\
     &$M_{\rm max}=1.97\,(2.21)$  & $M_{\rm max}=2.09\,(2.38)$  \\
1.00 &R=10.69 km                  & R=11.80 km                  \\
     &$\rho_c=7.23 \,\rho_0$      & $\rho_c=6.17 \,\rho_0$      \\
%    &                   &                                        \\
\hline
%    &                   &                                        \\
     &$M_{\rm max}=1.98\,(2.23)$  & $M_{\rm max}=2.15\,(2.47)$  \\
1.03 &R=10.81 km                  & R=12.39 km                  \\
     &$\rho_c=7.15 \,\rho_0$      & $\rho_c=5.75 \,\rho_0$      \\
%    &                   &                                        \\
\hline \hline
%    &   \multicolumn{2}{|c|}{}                                  \\
     & \multicolumn{2}{|c|}{$s=0, Y_{\nu_e}=0$}                 \\
%    & \multicolumn{2}{|c|}{}                                     \\
\hline \hline
%    &     \multicolumn{2}{|c|}{}                                  \\
     & \multicolumn{2}{|c|}{$M_{\rm max}=2.05\,(2.39) $}        \\
1.00 & \multicolumn{2}{|c|}{R=11.11 km}                         \\
     & \multicolumn{2}{|c|}{$\rho_c=6.91 \, \rho_0$}            \\
%    &     \multicolumn{2}{|c|}{}                              \\
\hline \hline\hline
    &       \multicolumn{2}{|c|}{}                            \\
        & \multicolumn{2}{|c|}{$npH$}                          \\
    &     \multicolumn{2}{|c|}{}                            \\
\hline \hline
%        &                            &                          \\
$q$     & $s=1$, $Y_{L}=0.4$         & $s=2$, $Y_{\nu_e}=0$     \\
%        &                            &                          \\
\hline \hline
%    &                   &                                        \\
                    & $M_{\rm max}=1.75\, (1.93)$    & $M_{\rm max}=1.55\, (1.69)$     \\
0.97                & R=11.11 km                     & R=12.39 km                      \\
                    & $\rho_c=6.84 \,\rho_0$         & $\rho_c=5.59 \,\rho_0$          \\
%    &                   &                                        \\
\hline
%    &                   &                                        \\
                    & $M_{\rm max}=1.76\,(1.94) $    & $M_{\rm max}=1.59\,(1.74)$      \\
1.00                & R=11.41 km                     & R=12.75 km                      \\
                    & $\rho_c=6.62 \,\rho_0$         & $\rho_c=5.44 \,\rho_0$          \\
%    &                   &                                        \\
\hline
%    &                   &                                        \\
                    & $M_{\rm max}=1.78\,(1.96)$     & $M_{\rm max}=1.71\,(1.90)$      \\
1.03                & R=11.56 km                     & R=13.34 km                      \\
                    & $\rho_c=6.51 \,\rho_0$         & $\rho_c=5.05 \,\rho_0$          \\
%    &                   &                                        \\
\hline \hline
%    &     \multicolumn{2}{|c|}{}                     \\
                 & \multicolumn{2}{|c|}{$s=0, Y_{\nu_e}=0$}                         \\
%    &      \multicolumn{2}{|c|}{}                           \\
\hline \hline
%    &         \multicolumn{2}{|c|}{}                          \\
                    & \multicolumn{2}{|c|}{$M_{\rm max}=1.57\, (1.76)$}               \\
1.00                & \multicolumn{2}{|c|}{R=12.35 km}                                \\
                    & \multicolumn{2}{|c|}{$\rho_c=5.66 \, \rho_0$}                   \\
%    &           \multicolumn{2}{|c|}{}                     \\
\hline \hline
\end{tabular}
\end{table}
%
%

%\newpage

\section{Conclusions}
\label{conclusion}

We have investigated the physical properties of the PNS in the framework of a
relativistic mean field theory based on nonextensive statistical
mechanics, characterized by power-law quantum distributions. We
have studied the finite temperature EOS in $\beta$-stable matter in absence and in presence of hyperons and trapped neutrinos. From a phenomenological point of view, we have
considered the nonextensive index $q$ as a free parameter, even
if, in principle, it should depend on the physical conditions
inside the PNS, on the fluctuation of the temperature and be
related to microscopic quantities (such as, for example, the mean
interparticle interaction length).
In this context, let us remember that, in the diffusional approximation, a value
$q\ne 1$ implies the presence of an anomalous diffusion among the constituent particles (the mean
square displacement obeys to a power-law behavior $<x^2>\propto t^\alpha$ with $\alpha\ne 1$).

We have shown that nonextensive statistical effects could play a crucial role in the structure and
in the evolution of the PNS also for small deviations from the standard Boltzmann-Gibbs statistics. As
expected, nonextensive statistical effects result to be particularly important during the maximum heating phase ($s=2, Y_{\nu_e}=0$), while become less relevant during the initial lepton rich state ($s=1, Y_{L}=0.4$) and negligible in the cold catalyzed phase ($s=0, Y_{\nu_e}=0$), due to the low temperatures achieved.

We have studied the relevance of nonextensive statistical effects:  i) in the temperature behavior  as a function of the baryon density, ii) in the softening of the EOS and, consequently, in the central baryon densities reached in the PNS core at fixed baryon mass, iii) in the lepton chemical potentials, iv) in the particle concentrations, in the hyperons formation and in the strangeness per baryon at fixed total baryon mass, v) in the maximum gravitational and baryonic masses.
Such a variation of physical quantities, respect to the standard case, can imply important consequences on the determination of the matter specific heat, on the neutrino mean free path inside the stellar core and, consequently, in the neutrino opacity and luminosity.

We have considered both cases of sub-extensive ($q<1$) and super-extensive ($q>1$) statistical effects which entail a sensible difference on the power-law particle distribution in the high energy region. When the entropic $q$ parameter is smaller than one, the energy tail of the particle distribution is depleted, otherwise, when $q$ is greater than one, the energy tail is enhanced. In the PNS context, the physical meaning of the difference between a sub-extensive or a super-extensive statistical behavior is reflected in different and well distinguishable phenomenological PNS properties.
For $q<1$, we have a reduction in temperature at fixed baryon density with respect to the standard case ($q=1$). Especially in the maximum heating phase, the EOS becomes slightly softer and higher central baryon densities at fixed total baryon mass are reached, influencing the neutrino diffusion during the deleptonization process. In the case of sub-extensive effects, hyperons start later but have a bigger concentration at high baryon density, allowing to sustain a higher neutrino luminosity at late times.
The other way round takes place in the case of $q>1$. We have an increase of the temperature as a function of the baryon density and lower central densities at fixed baryon mass are reached. The hyperons on-set is shifted at lower baryon densities and a greater hyperons concentration at low baryon masses is present. On the other hand, a significant reduction of the hyperons concentration at high stellar masses take place, contributing to a lower luminosity at late times. Moreover, we have shown that, in presence of super-extensive statistical effects and hyperon degrees of freedom, it is favored the realization of a metastable phase, with an enhancement of a possible black hole formation after the deleptonization era.

%\newpage

\end{document}